\documentclass[useAMS,usenatbib]{mn2e}
\usepackage[dvips]{graphicx}
\usepackage{amssymb}
\usepackage{longtable}

\def\aj{AJ}%
\def\apj{ApJ}%
\def\apjl{ApJ}%
\def\apjs{ApJS}%
\def\aap{A\&A}%
\def\mnras{MNRAS}%
\def\pasp{PASP}%
\title[The Colours of Elliptical Galaxies]
{The Colours of Elliptical Galaxies}
\author[R. Chang et al.]{Ruixiang Chang$^{1}$$^,$$^{2}$, Anna Gallazzi$^{2}$,Guinevere Kauffmann$^{2}$,
St\'ephane Charlot$^{2}$$^,$$^{3}$,\newauthor{\v{Z}eljko Ivezi\'{c}$^{4}$,
Jarle Brinchmann$^{5}$,
Timothy M. Heckman$^{6}$}\\
$^{1}$Shanghai Astronomical Observatory, 80 Nandan Road, Shanghai, China,
200030\\
$^{2}$Max Planck Institut fur Astrophysik, D-85748 Garching,
Germany\\
$^{3}$Institut d'Astrophysique du CNRS, 98 bis Boulevard Arago,
F-75014 Paris, France\\
$^{4}$Astronomy Department, University of Washington, Seattle, WA
98195, USA\\
$^{5}$Centro de Astrofisica da Universidade do Porto, Rua das
Estrelas, 4150-762 Porto,Portugal\\
$^{6}$Department of Physics and Astronomy, Johns Hopkins University,
Baltimore, MD 21218 USA\\}
\begin{document}

\date{Accepted . Received ; in original form...}
\pagerange{\pageref{firstpage}--\pageref{lastpage}} \pubyear{2005}

\maketitle
\label{firstpage}

\begin{abstract}
We have compiled a sample of 2728 nearby ($z<0.08$) elliptical
galaxies with photometry in the $g,r,i,z$ bands from the Sloan
Digital Sky Survey (SDSS) and $J,H,K$ photometry from the Two
Micron All Sky Survey (2MASS). Stellar masses, stellar velocity
dispersions and structural parameters such as sizes and surface
mass densities are also available for these objects. In order to
correct the aperture mismatch between SDSS and 2MASS, we correct
the SDSS magnitudes to the isophotal circular radius where the
2MASS magnitudes are measured. We compare the correlations between
optical, optical-infrared and infrared colours and galaxy
luminosity, stellar mass, velocity dispersion and surface
 mass density. We find that all galaxy colours correlate more strongly with
 stellar mass and velocity dispersion than with any other structural
 parameter. The dispersion about these two relations is also smaller.
 We also study the correlations between a variety of stellar absorption line
 indices and the same set of galaxy parameters and we reach very similar
 conclusions. Finally, we analyze correlations between absorption line
 indices and colour. Our results suggest that the optical colours of elliptical
 galaxies are sensitive to a combination of age, metallicity and
$\alpha$-enhancement, while the optical-infrared colours are
sensitive to metallicity and to $\alpha$-enhancement, but are
somewhat less sensitive to age.
\end{abstract}

\section{Introduction}
It is well known that the observed properties of elliptical
galaxies obey  a number of tight relations: the most well-studied
of these are the colour-magnitude relation (CMR), the fundamental
plane and the Mg-$\sigma$ relation
\citep{FJ76,vs77,dressler87,DD87,ble92,ellis97,bernardi03a,bernardi03b,bernardi03c,bernardi03d}.
These correlations link the properties of the stellar populations
of early-type galaxies with their masses and their structural
properties and are believed to encode important information about
how early-type galaxies formed.

It has long been known that early-type galaxies in nearby clusters
exhibit a tight colour-magnitude relation, i.e. more luminous
ellipticals  have redder colours than less luminous ellipticals
\citep{vs77,ble92}. This work has been
extended to moderate redshift and it is  found that there is no
significant change in the slope with redshift
\citep{aragon93,stanford95,stanford98,ellis97,kodama98,holden04}. The
zero-point of the relation shows
modest colour evolution,  consistent with the passive ageing of a
stellar population formed before $z \simeq 2$. Furthermore, the
slope of the CMR of ellipticals has been found to show little
dependence on environment \citep{terlevich01,bernardi03d,hogg04}.

The origin of the CMR is still a matter of controversy. The
conventional interpretation of the CMR is that it is primarily a
metallicity effect \citep{faber77,dressler84,ay87,kodama97,fcs99}.
It is often assumed that elliptical galaxies form monolithically
in a single giant burst of star formation at high redshift. Since
the binding energy per unit mass of gas is higher in more massive
galaxies, they are able to retain their gas for  longer and so
reach higher metallicities than less massive galaxies
\citep{larson74,ay87,bressan96,kodama97}. \cite{kc98} and
\cite{delucia04} have argued that the CMR can also be explained in
hierarchical models of galaxy formation when strong feedback is
included. In these models, the CMR for elliptical galaxies arises
because large ellipticals form by mergers of massive, metal-rich
progenitor disk systems. Indeed, some evidence of that build-up in
the early-type galaxy population has been observed
\citep{chen03,Bell04,cross04,fsw05}. However, it is also possible
to explain at least part of the CMR as an increase in mean stellar
age with luminosity \citep{worthey96,fcs99}. Furthermore,
\cite{jorgensen99} and \cite{trager00} find strong evidence for a
significant intermediate-age population in some elliptical
galaxies.

Additional information about stellar populations can be gained  by studying the
strength of stellar absorption features in galaxy spectra. The standard way of
quantifying absorption line strengths in galaxies is via the Lick-IDS system,
which includes 25 spectral absorption features defined by  a  `feature' bandpass
and two `pseudo-continuum' bandpasses. The indices are calibrated using the
spectra of 460 Galactic stars over the wavelength range from 4000\AA\ to 6400\AA.
The sensitivities of different Lick indices to stellar age and metallicity have
been discussed in a number of papers
\citep[][]{burstein84,gorgas93,trager98,henry99}. The H-Balmer lines are primarily
age-sensitive. Likewise, the 4000\AA~ break (i.e. the ratio of the flux blueward
and redward of 4000\AA) gets stronger with age, but for old stellar populations,
it shows also a second-order sensitivity to metallicity. Indices which are
primarily sensitive to metallicity include Fe and Mg lines between 4500\AA~ and
5700\AA~ and  molecular features such as the TiO band. In massive elliptical
galaxies, the abundance of $\alpha$-elements with respect to Fe-peak elements can
differ from the scaled-solar abundance ratio \citep[][]{wfg92}. Indices which
trace the abundance of either $\alpha$-elements or Fe-peak elements can therefore
be sensitive to the degree of $\alpha$-enhancement \citep[][]{Thomas03a}.

In addition to correlations with galaxy luminosity, past studies
of elliptical galaxies have also focused on how colours and
absorption line indices depend on the stellar velocity dispersion
of the galaxy. To first order, both the luminosity and the
velocity dispersion of an elliptical are measures of its mass; the
luminosity  measures the mass contained in stars and the velocity
dispersion is a dynamical probe of the total  mass contained
within some radius. The star formation history of a galaxy,
however, may not depend primarily  on its  mass. Indeed,
\cite{kennicutt98} has shown that star formation rates in spiral
galaxies correlate best with the local surface density of gas and
a law of the form $\Sigma_{SFR}=(2.5 \pm 0.7) \times
(\frac{\Sigma_{gas}}{1M_{\odot}pc^{-2}})^{1.4 \pm 0.15}
M_{\odot}yr^{-1}kpc^{-2} $  provides a good fit to the data.
\cite{BdJ00} studied trends in the optical and infrared colours of
a sample of nearby spiral galaxies and found that the colours most
sensitive to star formation history correlate best with stellar
surface density. Furthermore, \cite{kauf03b} studied the relations
between stellar mass, star formation history, size and internal
structure of a complete sample of $10^5$ galaxies and found that
the $\mu_\ast - M_\ast$ relation for the late-type galaxies can be
well described by a single power law. They also found that the
star formation history of low-mass galaxies are correlated more
strongly with surface density than stellar mass. These results
suggest that local rather than global factors regulate the rate at
which spiral galaxies form their stars at the present day.
Motivated by the surface density dependence of spiral star
formation histories, it seems worthwhile to search for surface
density dependence in the properties of early-type galaxies.

Finally, in a recent work \cite{eisenstein03} studied how the
average optical spectra of massive galaxies in the SDSS vary as a
function of both luminosity and environment. They found that as
redshift, luminosity and environment change, the variation of
pairs of Lick Indices appears to follow a simple one-dimensional
locus, suggesting that variations in these three parameters do not
lead to independent changes in the spectra of elliptical galaxies.

In this paper, we present the correlations between the colours of elliptical
galaxies and  a range of different {\em structural parameters}. We make use of
data from the Sloan Digital Sky Survey (SDSS) and the Two Micron All Sky Survey
(2MASS) to construct a sample of early-type  galaxies with photometry in five
optical ($u,g,r,i,z$) and  three near-infrared bands \it J \rm (1.25$\mu$m), \it H
\rm (1.65$\mu$m), \it $K_s$ \rm (2.17$\mu$m) bands. Stellar velocity dispersions
and size measurements for these systems are available from the SDSS database and
estimates of stellar mass are taken from the work of \cite{kauf03a}. The outline
of the paper is as follows. Section 2 describes how our sample is selected, the
method adopted for doing the aperture correction and calculating (K+E)-correction
of magnitudes. Section 3.1 presents the correlations between optical,
optical-infrared, infrared colours and structural parameters. In section 3.2, we
study the relations between colours and a
variety of Lick indices. Finally, in  section 4 we summarize our findings and give
our conclusions.

\section{The sample}

\subsection{Sample selection}

The SDSS is an imaging and spectroscopic survey of the high
Galactic latitude sky visible from the Northern hemisphere, which
will obtain \it u, g, r, i \rm and \it z \rm photometry of almost
a quarter of the sky and spectra of at least 700,000 objects
\citep{york}. A description of the software and data products are
given in \cite{stoughton} and \cite{dr103,dr204}. The 2MASS is a
ground-based, near-infrared imaging survey of the whole sky and
its extended source catalog (XSC) contains almost 1.6 million
galaxies \citep{jarrett00a,jarrett00b}. We positionally match
sources observed by the 2MASS to galaxies in the `main'
spectroscopic sample of the SDSS Data Release One (DR1; Abazajian
et al. 2003). Galaxies in this sample have $14.5 < r < 17.77$.
Practically all 2MASS sources in the DR1 area ($\sim 97$\% for
extended sources from the XSC) are matched to an SDSS counterpart
within 2 arcsec. We restrict our analysis to a  narrow range of
redshift $0.02<z<0.08$ in order to minimize uncertainties
introduced by K-corrections. The entire matched sample contains
8166 galaxies in this range of redshift.

In this paper, we focus on early-type  galaxies. To select these
systems, we require that they have 4000 \AA\ break strength
greater than 1.6 \citep[we adopt the narrow definition of the
break given in][]{balogh99} and concentration parameter $ C =
R_{90,r}/R_{50,r}> 2.6$ (where $R_{90,r}$ and $R_{50,r}$ are the
radii enclosing 90\% and 50\% of the total Petrosian light of the
galaxy). \cite{kauf03b} have shown that this cut separates
massive, dense, early-type galaxies from low mass, low density
late-type systems. In order to exclude systems with ongoing star
formation or residual AGN activity, we also eliminate all galaxies
where the equivalent width of the  H$\alpha$ emission line is
greater than 2 \AA\ . Our final sample contains 2728 early-type
galaxies.

\subsection{Photometric quantities and aperture corrections}

In order to compare optical and optical-infrared colours, it is
important that both SDSS and 2MASS magnitudes are measured within
the same aperture.
There are several sets of magnitudes provided in the 2MASS XSC:
elliptical isophotal magnitudes, circular isophotal
magnitudes, total magnitudes using a Kron elliptical aperture, total
magnitudes from extrapolating the fit to the radial profile
and  magnitudes within
fixed circular apertures (5, 7, 10, 15 arcsec ). In this paper, we
use the isophotal fiducial magnitudes, which
are measured within the circular aperture corresponding to
a surface brightness of  20.0
mag arcsec$^{-2}$ in the $K_s$-band (the aperture is denoted as
$R_{k20fc}$). At this isophotal magnitude,
the background noise is still relatively low, but the aperture
is large enough to enclose most of the light from the galaxy
(see the 4th panel of Figure 1).
This set of magnitudes are measured
with the same size aperture in all three ($J,H,K_s$) 2MASS bands and
they thus provide a consistent set of infrared colours.

SDSS provides the azimuthally averaged surface brightness in a
series of circular annuli.
In order to derive optical-infrared colours, we have calculated  $g, r, i, z$
magnitudes that are matched to the 2MASS measurements
by interpolating the cumulative radial surface
brightness profile in each SDSS band at the corresponding
isophotal radius $R_{k20fc}$. From now on, we refer to
these corrected magnitudes as the ``SDSS magnitudes'' and they
are used to estimate the optical and optical-infrared colours
throughout the paper.

\begin{figure}
\centerline{\includegraphics[width=8truecm]{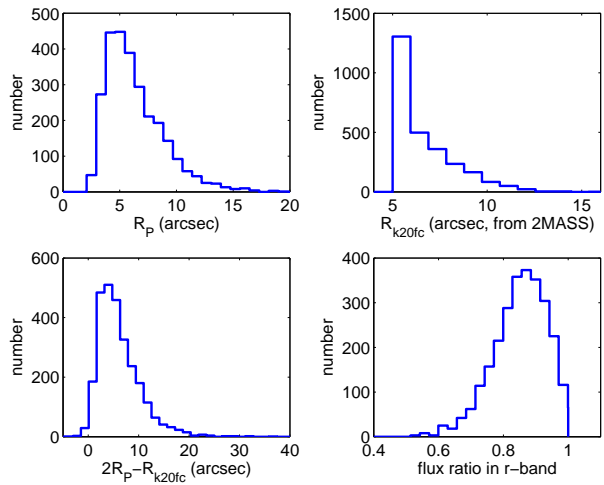}}
\caption{The distributions of the Petrosian radius in r-band $R_P$
(from SDSS), the isophotal radius $R_{k20fc}$ (from 2MASS), the
quantity $2R_P-R_{k20fc}$ and the $r$-band flux within $R_{k20fc}$
divided by the Petrosian flux (i.e., within 2$R_P$). }\label{fig1}
\end{figure}

In Fig.~\ref{fig1} we plot the distribution of the
$r$-band  Petrosian radius (from SDSS),
the isophotal radius $R_{k20fc}$ (from
2MASS), the quantity  $2R_P-R_{k20fc}$, and the $r$-band
flux within $R_{k20fc}$ divided by the Petrosian flux.
A small fraction of galaxies (about 1.2
percent) have isophotal radius larger than $2R_P$. We simply adopt the
Petrosian magnitudes for these galaxies and make no aperture
correction. Fig.~\ref{fig1} shows that for most
galaxies, a large fraction of
Petrosian flux is contained within  $R_{k20fc}$.
Because colour gradients are  small for early-type
galaxies \citep{wu05}, it is safe to treat the colours deduced
from the corrected SDSS magnitudes as a global galaxy colours.

Stellar mass estimates for this sample are available from the work of
\cite{kauf03a}. To estimate stellar masses, two stellar absorption line indices,
the 4000\AA\ break strength $D_n(4000)$ and the Balmer absorption line index
H$\delta_A$, are used to constrain the mean stellar age of a galaxy and the
fraction of its stellar mass formed in bursts over the past few Gyr. A comparison
with broad band magnitudes then yields estimates of dust attenuation and of
stellar mass. The optical photometric quantities, such as Petrosian magnitudes in
5 bands, size $R_{50,z}$, concentration parameter $C=R_{90,r}/R_{50,r}$, are
obtained from the First Data Release (DR1) of the SDSS \citep{dr103}. Velocity
dispersion measurements for each galaxy are also available from the database and
are described in Schlegel et al (in preparation).

Throughout the paper, we  assume a cosmology
with $\Omega_M=0.3$, $\Omega_{\Lambda}=0.7$ and $H_0$=70 km s$
^{-1}$ Mpc$^{-1}$.

\begin{table*}
\caption{The results of linear fits. The first set of quantities
are the best-fit slopes. The second set of quantities $disp$ give
the dispersions about the relation, which we  define as
$\sum_{i=1}^n\frac{|y_i^o-y_i^m|}{n}$, where $y_i^o$ is the
observed colour  and $y_i^m$ is the colour  predicted by the fit
and $n$ is the number of galaxies. The third and fourth set of
quantities $\Delta I$ measure the changes in colour over the
interval in magnitude, mass, $\sigma$ or density that contains
$90\%$ of the galaxies (we exclude the lower and upper 5th
percentiles of the distribution). $\Delta I$ are normalized by
dividing either by the dispersion (set 3) or by the total range in
colour enclosing 90\% of the galaxies (set 4). } \centering
\begin{tabular}{lccccccc}
\\
\hline \hline & $M_r$ & $ \log(M_\ast) $ & $ \log(\sigma_{meas}) $
& $
\log(\sigma_{inf}) $ &  $ \log(\mu_\ast) $ & $ \log(\rho_\ast) $\\

\hline & \\
slope (g-r) & -0.0247 & 0.0681  &  0.1869  &  0.1957  &  0.0635 &   0.0131 \\
slope (r-i) & -0.0115 & 0.0378  &  0.0913  &  0.1030  &  0.0341 &   0.0081  \\
slope (r-z) & -0.0216 &  0.0732 &   0.1910 &   0.2111 &   0.0809 &   0.0239  \\
slope (r-J) & -0.0488 &  0.1392 &   0.3521 &   0.4194 &   0.1442 &   0.0335  \\
slope (r-K) & -0.0731 &  0.1885 &   0.4586  &  0.5613  &  0.1683  &  0.0294 \\
slope (J-K) & -0.0182 & 0.0553  &  0.1202   & 0.1347  &  0.0259  & -0.0053 \\

\\
disp (g-r) &  0.0330 &  0.0295 &   0.0292  &  0.0294  &  0.0334  &  0.0359  \\
disp (r-i) & 0.0218  & 0.0199  &  0.0201 &   0.0198  &  0.0216  &  0.0228  \\
disp (r-z) &  0.0395 & 0.0354  &  0.0352 &   0.0345  &  0.0378  &  0.0406   \\
disp (r-J) & 0.0783 & 0.0724 &   0.0738  &  0.0712  &  0.0781  &  0.0829  \\
disp (r-K) & 0.1090 & 0.1006 &   0.1043  &  0.1008  &  0.1110  &  0.1160  \\
disp (J-K) & 0.0748 & 0.0739 &   0.0743  &  0.0743  &  0.0752  &  0.0754 \\

\\
$\frac {\Delta I}{disp}(g-r) $ & 1.6415 &   2.6352  &  2.6681  &  2.6968  &  1.4901 &   0.3832    \\
$\frac {\Delta I}{disp}(r-i) $ & 1.1583 &   2.1710  &  1.8933  &  2.1121  &  1.2357 &   0.3730  \\
$\frac {\Delta I}{disp}(r-z) $ & 1.1963 &   2.3589  &  2.2635  &  2.4747  &  1.6759 &   0.6165 \\
$\frac {\Delta I}{disp}(r-J) $ & 1.3642 &   2.1905  &  1.9905  &  2.3839  &  1.4465 &   0.4235 \\
$\frac {\Delta I}{disp}(r-K) $ & 1.4701 &   2.1357  &  1.8351  &  2.2543  &  1.1886 &   0.2655 \\
$\frac {\Delta I}{disp}(J-K) $ & 0.5338 &   0.8542  &  0.6748  &  0.7345  &  0.2701 &  -0.0732 \\
\\
$\frac {\Delta I}{range}(g-r) $ & 0.3509 &   0.5021  &  0.5045 &   0.5125 &   0.3223  &  0.0889 \\
$\frac {\Delta I}{range}(r-i) $ & 0.2675 &   0.4566  &  0.4038 &   0.4419 &   0.2830  &  0.0899 \\
$\frac {\Delta I}{range}(r-z) $ & 0.2680 &   0.4730  &  0.4518 &   0.4842 &   0.3592  &  0.1418 \\
$\frac {\Delta I}{range}(r-J) $ & 0.2961 &   0.4395  &  0.4071 &   0.4704 &   0.3131  &  0.0973 \\
$\frac {\Delta I}{range}(r-K) $ & 0.3202 &   0.4294  &  0.3825 &   0.4542 &   0.2637  &  0.0615 \\
$\frac {\Delta I}{range}(J-K) $ & 0.1177 &   0.1861  &  0.1479 &   0.1609 &   0.0599 &  -0.0163 \\

\hline\hline
\end{tabular}
\end{table*}

\begin{figure*}
\centerline{\includegraphics[width=10truecm]{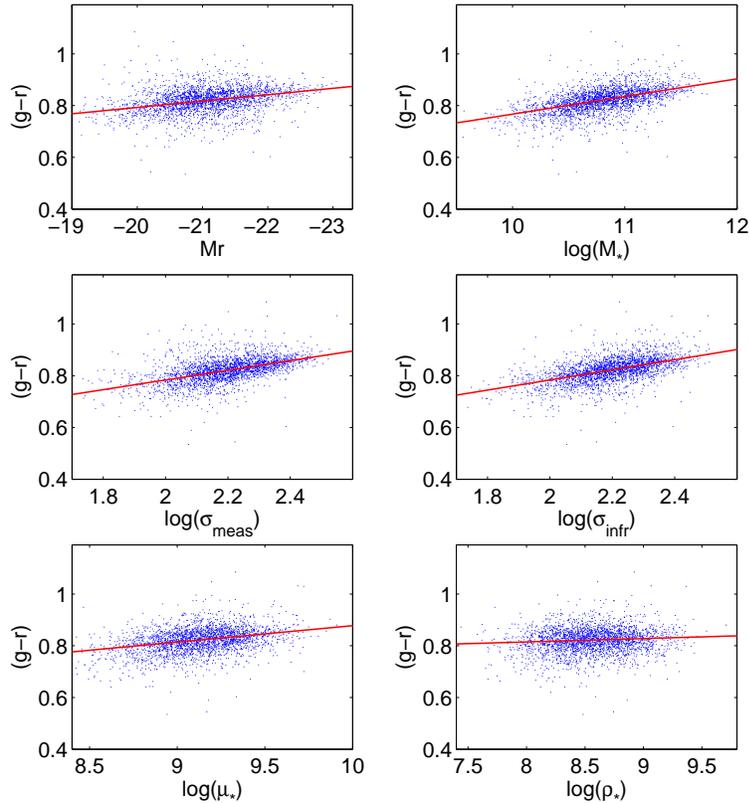}}
\caption{The correlations between $g-r$ and $r$-band magnitude
$M_r$, stellar mass $M_\ast$ (in units of M$_{\odot}$), measured
stellar velocity dispersion $\sigma_{meas}$ (in units of km/s),
``inferred'' stellar velocity dispersion $\sigma_{inf}=0.5
(GM_\ast/R_{50,z})^{1/2}$ (in units of km/s), stellar surface mass
density $\mu_*= 0.5 M_\ast/(\pi R_{50,z}^2)$ (in unit of
M$_{\odot}$/kpc$^2$) and density $\rho_\ast= 0.5 M_\ast /(4/3 \pi
R_{50,z}^3)$ (in units of M$_{\odot}$/kpc$^3$). The solid lines
show linear fits using robust techniques. }\label{fig2}
\end{figure*}

\begin{figure*}
\centerline{\includegraphics[width=10truecm]{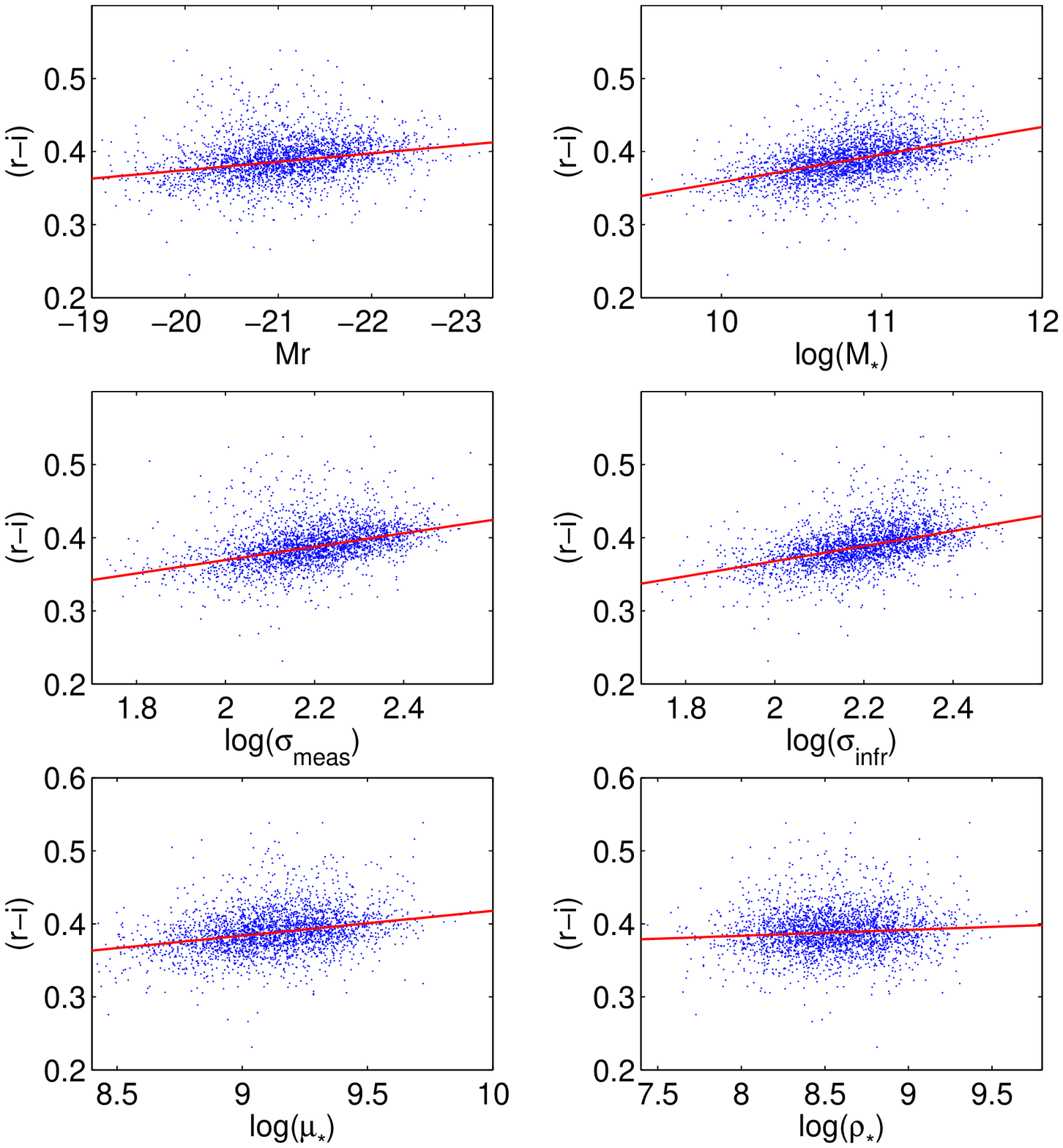}}
\caption{Same as figure 2, but for $r-i$ colour.}\label{fig3}
\end{figure*}

\begin{figure*}
\centerline{\includegraphics[width=10truecm]{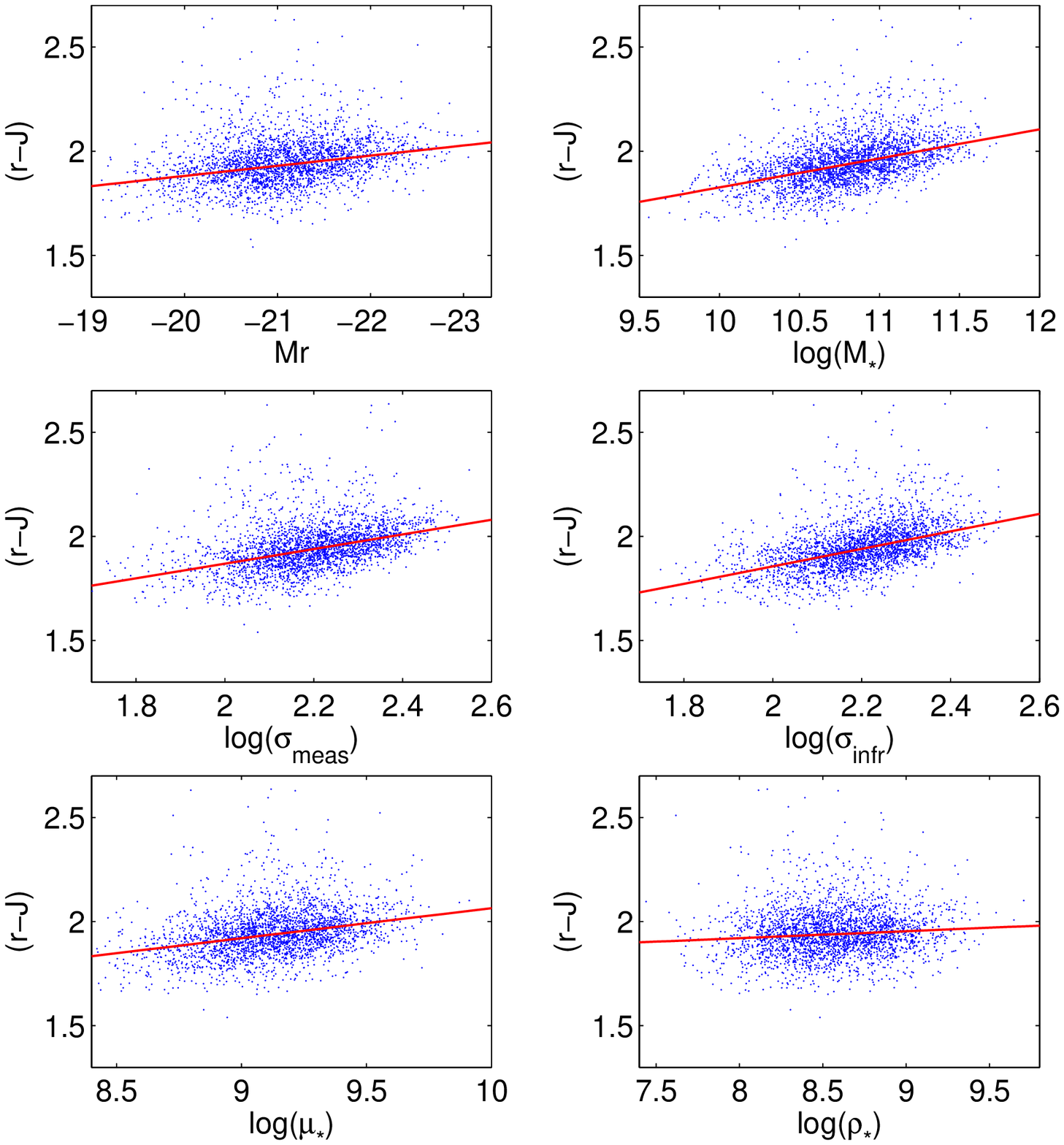}}
\caption{Same as figure 2, but for $r-J$ colour.}\label{fig4}
\end{figure*}

\begin{figure*}
\centerline{\includegraphics[width=10truecm]{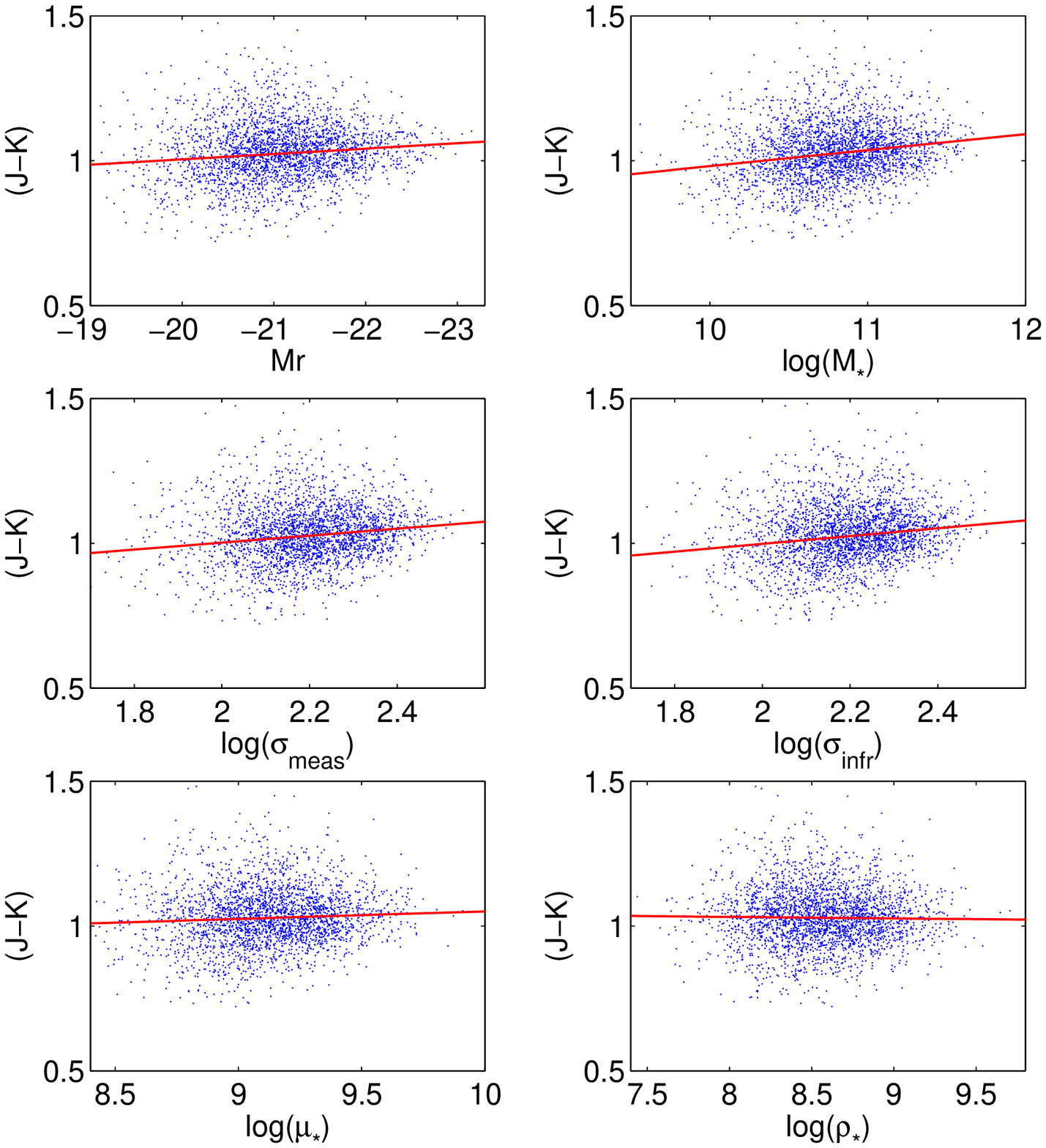}}
\caption{Same as figure 2, but for $J-K$ colour.}\label{fig5}
\end{figure*}

\subsection{(K+E)-correction}

When converting from the  apparent magnitude to the
rest-frame absolute magnitude of an object, the fact that the SDSS
and 2MASS filters measure the light from a fixed spectral range
in the local frame needs to  be taken into account. Correcting for this
effect is known as the K-correction. The E-correction accounts for
the change in the galaxy's luminosity between the time that its
light was emitted and the present day. In this paper, the sum of
K-correction and E-correction is denoted as the (K+E)-correction.

In order to evaluate these corrections, we generate a grid of
model spectral energy distributions (SEDs) using the population
synthesis models of \cite[][BC03]{bc03}. We adopt the
universal initial mass function  from \cite{chab03} and  lower
and upper mass cut-offs are taken as $m_L$=0.1M$_{\odot}$ and
$m_U$=100M$_{\odot}$, respectively. We parameterize the models
according to a metallicity $Z$ and the star formation timescale
$\tau$. Galaxies form stars according to the equation
\begin{equation}
\Psi(t)=\tau^{-1}e^{-t/\tau},
\end{equation}
for a total duration of 12 Gyr.  The effect of  extinction on the
SEDs is not included, because the dust  attenuation measurements
of \cite{kauf03a} show that there is very little dust in these
systems. For each model, we store observer frame colours at a
series of closely spaced redshifts from z=0.08 to z=0.02. For each
galaxy in our sample, we find the model that most closely
reproduces its observed colours (including $g-r$, $r-i$, $r-z$,
$r-J$, $r-K$ and $J-K$) and we use the model to transform to
colours and magnitudes defined at z=0.04. Because of the very
limited range of redshifts of the galaxies in our sample, these
colour corrections are small: 0.04 mag for $g-r$, 0.01 mag for
$r-i$ and $r-z$, 0.03 mag for $r-J$, 0.05 mag for $r-K$ and 0.02
mag for $J-K$ on average.

\section{Results}
\subsection{Relations between colours and structural parameters}

In this section, we present the correlations of the optical,
optical-infrared and infrared colours of the ellipticals in our
sample with the following parameters:
\begin{enumerate}
\item absolute $r$-band  magnitude, $M_r$ (Note that we use
Petrosian magnitudes to calculate $M_r$); \item stellar mass
$M_\ast$; \item measured stellar velocity dispersion
$\sigma_{meas}$; \item ``inferred'' stellar velocity dispersion:
$\sigma_{inf}=C (GM_\ast/R_{50,z})^{1/2}$, where $C$ was chosen so
that the average value of $\sigma_{inf}$ for all the galaxies in
the sample  would be equal to the average value of $\sigma$ for
the same set of galaxies (we find $C=0.5$); \item stellar surface
mass density $\mu_\ast= 0.5 M_\ast/(\pi R_{50,z}^2)$; \item
stellar mass ``density'' $\rho_\ast= 0.5 M_\ast /(4/3 \pi
R_{50,z}^3)$.
\end{enumerate}
where $R_{50,z}$ is  the radius enclosing 50 percent of Petrosian
flux in $z$-band (and hence approximately 50 percent of the
stellar mass). \footnote {We also tried  different measurements of
z-band half-light radius including the seeing deconvolved Sersic
and Petrosian half-light radii derived by \cite{blanton03}. We
obtained almost identical results.} In this paper, the parameters
$M_\ast$, $\sigma_{meas}$, $\sigma_{inf}$, $\mu_\ast$, $\rho_\ast$
are in units of M$_{\odot}$, km/s, km/s, M$_{\odot}$/kpc$^2$ and
M$_{\odot}$/kpc$^3$, respectively. The results are presented in
Figures 2-5, where we show the correlations between $g-r$, $r-i$,
$r-J$ and $J-K$ colours as a function of these parameters. We find
that a simple least-squares fit to the data is sensitive to
whether or not we exclude galaxies that lie far away from the main
relations. We thus switched to robust  techniques and we fit a
straight line to the data by minimizing the absolute deviations.
This is shown as a solid line on the plots.

Table 1 lists the parameters of  the fits. The first set of
quantities are the best-fit slopes. The second set of quantities
$disp$ give the dispersions about the relation, which we  define
as $\sum_{i=1}^n\frac{|y_i^o-y_i^m|}{n}$, where $y_i^o$ is the
observed colour  and $y_i^m$ is the colour  predicted by the fit
and $n$ is the number of galaxies. The third and fourth set of
quantities $\Delta I$ measure the changes in colour over the
interval in magnitude, mass, $\sigma$ or density that contains
90\% of the galaxies (we exclude the lower and upper 5th
percentiles of the distribution). In order to compare these colour
changes between different photometric bands, we normalize $\Delta
I$  by dividing either by the dispersion ($\Delta I/disp$) or by
the total range in colour enclosing  90\% of the galaxies ($\Delta
I/range$).

We find that the optical ($g-r$, $r-i$, $r-z$), optical-infrared
($r-J$, $r-K$) and infrared ($J-K$) colours all correlate with
$r$-band absolute magnitude, i.e. more luminous ellipticals tend
to have redder colours than less luminous ellipticals. Table 1
shows that colours correlate more strongly with stellar mass and
velocity dispersion than with $r$-band magnitude and that the
dispersions about the relations are also smaller. Our results are
consistent with those of \cite{bernardi03d,bernardi05}, who
demonstrate that the colour-magnitude relation is a consequence of
the fact that both the luminosity and the colours of early-type
galaxies are correlated with their stellar velocity dispersions.
These results suggest that the colours of elliptical galaxies are
primarily determined by their {\em mass}.

\begin{table*}
\caption{Relations between index strengths (corrected for velocity
dispersion) and $g-r$ and $r-J$ colours, fitted with our robust
technique on the subsample of galaxies that lie within 2.5 times the
dispersion in the $g-r,M_r$ and $r-J,M_r$ colour-magnitude relations.
Column 2 gives the average S/N ratio of each index, calculated as the
ratio between the index strength and the observational error. For
each colour the first and second columns give the slope and rms
scatter of the relation. The third column indicates the correlation
coefficient according to a Spearman rank-order test. The fourth
column gives the change in index strength (along the fitted relation)
over the interval in colour that contains 90 percent of the points
($\Delta$ I), normalized to the dispersion about the fitted relation
(disp). Columns 11 and 12 provide some indications about the indices.
Column 11 lists the elements that are expected to contribute most to
the index strength (in the case of iron, a `+' indicates a weak
influence and a `++' a strong influence). Column 12 indicates the
sensitivity of the index to $\alpha$/Fe variations, as expected from
Henry \& Worthey (1999), Thomas et al. (2003) and Thomas et al
(2004). Up and down arrows, respectively, correspond to increasing
and decreasing index strengths with increasing $\alpha$/Fe ratio,
slanted arrows to intermediate trends, and a hyphen to no
sensitivity.}\label{idx_colour}
\centering
\begin{tabular}{|lc|rrrr|rrrr|cc|}
\noalign{\smallskip} \hline \noalign{\smallskip}
Index & Log(S/N)& \multicolumn{4}{c}{$g-r$} & \multicolumn{4}{c}{$r-J$} & & \\ 
 & & slope & disp & $\sigma_{corr}$ & $\rm \Delta I/disp$ & slope & disp & $\sigma_{corr}$ &
 $\rm \Delta I/disp$ & element & $\alpha/Fe$\\
 (1) & (2) & (3) & (4) & (5) & (6) & (7) & (8) & (9) & (10) & (11) & (12)\\
\noalign{\smallskip} \hline \noalign{\smallskip}
$CN_1$                   &   0.501 &   0.71 &  0.023 &  0.62 &  4.74 &   0.21  &  0.027  &   0.42 &   2.85  & C,N & $\uparrow$     \\
$CN_2$           &   0.794 &   0.75 &  0.026 &  0.60 &  4.50 &   0.22  &  0.030  &   0.41 &   2.69  & C,N & $\uparrow$     \\
Ca4227           &   0.757 &   2.01 &  0.221 &  0.24 &  1.41 &   0.74  &  0.225  &   0.18 &   1.18  & Ca,C,N &--       \\
G4300            &   1.171 &   4.08 &  0.405 &  0.26 &  1.56 &   0.85  &  0.417  &   0.15 &   0.74  & C,O,Fe & $\searrow$    \\
Fe4383           &   1.113 &   5.21 &  0.543 &  0.24 &  1.49 &   2.08  &  0.547  &   0.22 &   1.37  & ++Fe & $\downarrow$    \\
Ca4455           &   0.761 &   1.70 &  0.237 &  0.19 &  1.11 &   0.49  &  0.240  &   0.14 &   0.74  & Fe,Cr & --       \\
Fe4531           &   1.000 &   2.62 &  0.399 &  0.19 &  1.02 &   0.82  &  0.403  &   0.14 &   0.74  & +Fe & $\searrow$     \\
$C_2$            &   1.169 &  14.15 &  0.747 &  0.45 &  2.94 &   5.10  &  0.777  &   0.36 &   2.37  & C,N & $\searrow$     \\
$H\beta$         &   0.892 &  -4.29 &  0.264 & -0.39 & -2.52 &  -1.28  &  0.280  &  -0.24 &  -1.65  & --&  --          \\
Fe5015           &   1.137 &   4.15 &  0.507 &  0.21 &  1.27 &   1.68  &  0.508  &   0.20 &   1.19  & +Fe & $\searrow$     \\
$Mg_1$           &   1.240 &   0.42 &  0.014 &  0.62 &  4.81 &   0.14  &  0.016  &   0.48 &   3.20  & Mg,C & $\uparrow$    \\
$Mg_2$           &   1.470 &   0.57 &  0.020 &  0.61 &  4.52 &   0.20  &  0.022  &   0.50 &   3.28  & Mg & $\uparrow$      \\
Mgb          &   1.199 &   7.15 &  0.351 &  0.47 &  3.16 &   2.68  &  0.366  &   0.40 &   2.65  & Mg & $\uparrow$      \\
Fe5270           &   1.033 &   1.93 &  0.331 &  0.19 &  0.90 &   0.62  &  0.333  &   0.16 &   0.68  & ++Fe & $\searrow$    \\
Fe5335           &   0.993 &   2.68 &  0.324 &  0.25 &  1.28 &   0.99  &  0.326  &   0.21 &   1.09  & ++Fe & $\downarrow$    \\
Fe5406           &   0.865 &   1.70 &  0.250 &  0.18 &  1.05 &   0.73  &  0.250  &   0.18 &   1.05  & ++Fe & $\searrow$    \\
Fe5709           &   0.679 &   0.32 &  0.185 &  0.03 &  0.27 &   0.13  &  0.185  &   0.02 &   0.25  & +Fe & $\searrow$     \\
Fe5782           &   0.698 &   1.18 &  0.168 &  0.19 &  1.08 &   0.48  &  0.168  &   0.17 &   1.04  & +Fe & $\searrow$     \\
NaD          &   1.291 &  12.53 &  0.498 &  0.58 &  3.90 &   4.74  &  0.526  &   0.51 &   3.26  & Na, (ISM)& --    \\
$TiO_1$          &   0.758 &   0.03 &  0.007 &  0.13 &  0.71 &   0.01  &  0.007  &   0.13 &   0.75  & O,Ti& --         \\
$TiO_2$          &   1.273 &   0.09 &  0.008 &  0.32 &  1.80 &   0.04  &  0.008  &   0.34 &   2.01  & O,Ti& --         \\
$H\delta_A$          &   0.389 & -12.15 &  0.692 & -0.43 & -2.72 &  -3.57  &  0.741  &  -0.28 &  -1.74  & --& $\searrow$       \\
$H\gamma_A$          &   0.998 & -13.88 &  0.655 & -0.47 & -3.29 &  -4.24  &  0.711  &  -0.33 &  -2.16  & --& $\searrow$       \\
$D4000$                  &   1.950 &   1.74 &  0.060 &  0.60 &  4.53 &   0.51  &  0.070  &   0.39 &   2.65  & --& ?        \\
$[MgFe]^\prime$          &   1.311 &   4.41 &  0.252 &  0.44 &  2.71 &   1.53  &  0.264  &   0.38 &   2.10  & Mg,Fe& --        \\
$[Mg_1Fe]$           &   1.607 &   0.40 &  0.018 &  0.50 &  3.42 &   0.14  &  0.019  &   0.40 &   2.70  & Mg,Fe& --        \\
$[Mg_2Fe]$           &   1.662 &   0.48 &  0.021 &  0.52 &  3.54 &   0.18  &  0.022  &   0.44 &   2.89  & Mg,Fe& --        \\
$Mg_2/\langle Fe\rangle$ &   1.115 &   0.12 &  0.013 &  0.32 &  1.52 &   0.05  &  0.013  &   0.25 &   1.33  & Mg,Fe& $\uparrow$    \\
$Mgb/\langle Fe\rangle$  &   1.027 &   1.33 &  0.222 &  0.19 &  0.93 &   0.47  &  0.223  &   0.16 &   0.76  & Mg,Fe& $\uparrow$    \\
\noalign{\smallskip}
\hline
\noalign{\smallskip}
\end{tabular}
\end{table*}

\subsection{Stellar absorption features}\label{indices}
In this section we study the relations between stellar absorption features,
colours and structural parameters of the galaxies in our sample.

We have measured  Lick indices from the SDSS spectra, using the
bandpass definition of \cite{worthey97}. It should be noted that
the strengths of some of the absorption indices are strongly
affected by broadening due to the velocity dispersion $\sigma$ of
the stars in the galaxy. Indices measured with the narrowest
`pseudo-continuum' bandpasses show the strongest dependence on
$\sigma$. To properly compare index strengths of galaxies with
different velocity dispersion and study their intrinsic dependence
on various galaxy properties, we have to correct for this effect.
We normalize the index strengths of all galaxies to a common
velocity dispersion of $\rm 200~ km~ s^{-1}$. \footnote{The
average velocity dispersion of the galaxies in the sample is $\rm
160~ km~ s^{-1}$.} This is achieved by fitting a relation between
index strength and velocity dispersion for BC03 simple stellar
populations of different metallicities.  These relations are then
interpolated to the metallicity of the galaxy, as estimated using
the methods of \cite{gallazzi05}.

\begin{table*}
\caption{Relation between index strengths (corrected for velocity
dispersion) and structural parameters: stellar mass ($\log M_\ast$),
velocity dispersion ($\log \sigma_{meas}$) and surface stellar mass
density ($\log \mu_\ast$), fitted by applying our robust technique to
the subsample of galaxies that lie within 2.5 times the dispersion in
the $g-r,M_r$ and $r-J,M_r$ colour-magnitude relations. Columns
2,6,10 give the slope of the relation; columns 3,7,11 give the rms
scatter about the fitted relation; columns 4,8,12 give the
correlation coefficient according to a Spearman rank-order test.
Columns 5,9,13 give the change in index strength along the fitted
relation over the interval in structural parameter that contains 90
percent of the points, normalized to the dispersion about the fitted
relation.}\label{idx_par1}
\centering
\begin{tabular}{|l|rrrr|rrrr|rrrr|}
\noalign{\smallskip} \hline \noalign{\smallskip} Index &
\multicolumn{4}{c}{$\log M_\ast$} & \multicolumn{4}{c}{$\log
\sigma_{meas}$} &
\multicolumn{4}{c}{$\log \mu_\ast$}\\
 & slope & disp & $\sigma_{corr}$ & $\rm \Delta I/disp$ & slope & disp & $\sigma_{corr}$ &
 $\rm \Delta I/disp$ & slope & disp & $\sigma_{corr}$ & $\rm \Delta I/disp$\\
 (1) & (2) & (3) & (4) & (5) & (6) & (7) & (8) & (9) & (10) & (11) & (12) & (13)\\
\noalign{\smallskip} \hline \noalign{\smallskip}
$CN_1$                   &   0.06 & 0.024  &  0.56 &   3.02 &  0.21 &  0.02 &  0.70 &   4.24 &  0.060 &  0.03 &  0.32 &  1.66 \\
$CN_2$                   &   0.07 & 0.027  &  0.55 &   3.08 &  0.23 &  0.02 &  0.70 &   4.27 &  0.066 &  0.03 &  0.32 &  1.66 \\
Ca4227                   &   0.15 & 0.226  &  0.16 &   0.76 &  0.52 &  0.22 &  0.21 &   0.99 &  0.113 &  0.23 &  0.10 &  0.39 \\
G4300                    &   0.27 & 0.414  &  0.17 &   0.76 &  1.25 &  0.40 &  0.29 &   1.32 &  0.417 &  0.41 &  0.18 &  0.80 \\
Fe4383                   &   0.53 & 0.544  &  0.24 &   1.13 &  1.52 &  0.54 &  0.27 &   1.20 &  0.698 &  0.55 &  0.22 &  1.00 \\
Ca4455                   &   0.17 & 0.237  &  0.20 &   0.82 &  0.58 &  0.23 &  0.24 &   1.04 &  0.205 &  0.24 &  0.15 &  0.68 \\
Fe4531                   &   0.32 & 0.398  &  0.20 &   0.92 &  0.94 &  0.40 &  0.22 &   1.00 &  0.340 &  0.40 &  0.15 &  0.67 \\
$C_2$                    &   1.47 & 0.734  &  0.47 &   2.30 &  4.59 &  0.70 &  0.54 &   2.79 &  1.509 &  0.79 &  0.32 &  1.51 \\
$H\beta$                 &  -0.32 & 0.278  & -0.28 &  -1.31 & -1.36 &  0.25 & -0.46 &  -2.26 & -0.315 &  0.28 & -0.19 & -0.87 \\
Fe5015                   &   0.53 & 0.497  &  0.30 &   1.24 &  1.60 &  0.49 &  0.30 &   1.38 &  0.597 &  0.51 &  0.19 &  0.92 \\
$Mg_1$                   &   0.03 & 0.015  &  0.54 &   2.73 &  0.14 &  0.01 &  0.77 &   5.10 &  0.043 &  0.02 &  0.40 &  2.09 \\
$Mg_2$                   &   0.05 & 0.021  &  0.53 &   2.70 &  0.18 &  0.02 &  0.75 &   4.62 &  0.059 &  0.02 &  0.41 &  2.02 \\
Mgb                      &   0.59 & 0.366  &  0.39 &   1.86 &  2.46 &  0.32 &  0.59 &   3.25 &  0.889 &  0.37 &  0.37 &  1.91 \\
Fe5270                   &   0.20 & 0.331  &  0.18 &   0.71 &  0.59 &  0.33 &  0.19 &   0.76 &  0.196 &  0.33 &  0.14 &  0.46 \\
Fe5335                   &   0.29 & 0.321  &  0.26 &   1.04 &  0.91 &  0.32 &  0.29 &   1.21 &  0.294 &  0.33 &  0.18 &  0.71 \\
Fe5406                   &   0.18 & 0.250  &  0.18 &   0.84 &  0.56 &  0.25 &  0.22 &   0.97 &  0.190 &  0.25 &  0.14 &  0.59 \\
Fe5709                   &   0.02 & 0.185  &  0.03 &   0.11 & -0.07 &  0.18 & -0.03 &  -0.15 &  0.031 &  0.18 &  0.01 &  0.13 \\
Fe5782                   &   0.15 & 0.167  &  0.21 &   1.04 &  0.42 &  0.17 &  0.22 &   1.06 &  0.153 &  0.17 &  0.16 &  0.71 \\
NaD                      &   1.19 & 0.510  &  0.53 &   2.70 &  3.94 &  0.45 &  0.66 &   3.67 &  1.420 &  0.54 &  0.45 &  2.05 \\
$TiO_1$                  &   0.00 & 0.007  &  0.10 &   0.45 &  0.01 &  0.01 &  0.20 &   0.99 &  0.008 &  0.01 &  0.21 &  0.95 \\
$TiO_2$                  &   0.01 & 0.008  &  0.26 &   1.14 &  0.03 &  0.01 &  0.41 &   1.82 &  0.013 &  0.01 &  0.31 &  1.38 \\
$H\delta_A$              &  -0.95 & 0.730  & -0.32 &  -1.49 & -3.65 &  0.68 & -0.46 &  -2.25 & -1.115 &  0.75 & -0.24 & -1.17 \\
$H\gamma_A$              &  -1.18 & 0.691  & -0.38 &  -1.97 & -3.91 &  0.64 & -0.52 &  -2.60 & -1.172 &  0.73 & -0.26 & -1.27 \\
$D_n(4000)$              &   0.15 & 0.065  &  0.50 &   2.63 &  0.47 &  0.06 &  0.63 &   3.42 &  0.129 &  0.07 &  0.30 &  1.40 \\
$[MgFe]^\prime$          &   0.39 & 0.259  &  0.39 &   1.75 &  1.41 &  0.24 &  0.53 &   2.50 &  0.516 &  0.26 &  0.33 &  1.53 \\
$[Mg_1Fe]$               &   0.04 & 0.018  &  0.51 &   2.70 &  0.14 &  0.02 &  0.63 &   3.67 &  0.043 &  0.02 &  0.36 &  1.73 \\
$[Mg_2Fe]$               &   0.05 & 0.021  &  0.53 &   2.78 &  0.16 &  0.02 &  0.66 &   3.78 &  0.056 &  0.02 &  0.38 &  1.94 \\
$Mg_2/\langle Fe\rangle$ &   0.01 & 0.013  &  0.25 &   0.95 &  0.04 &  0.01 &  0.41 &   1.40 &  0.013 &  0.01 &  0.19 &  0.81 \\
$Mgb/\langle Fe\rangle$  &   0.08 & 0.224  &  0.12 &   0.39 &  0.49 &  0.22 &  0.26 &   0.95 &  0.172 &  0.22 &  0.16 &  0.61 \\
\noalign{\smallskip} \hline \noalign{\smallskip}
\end{tabular}
\end{table*}

\begin{figure*}
\centerline{\includegraphics[width=10truecm]{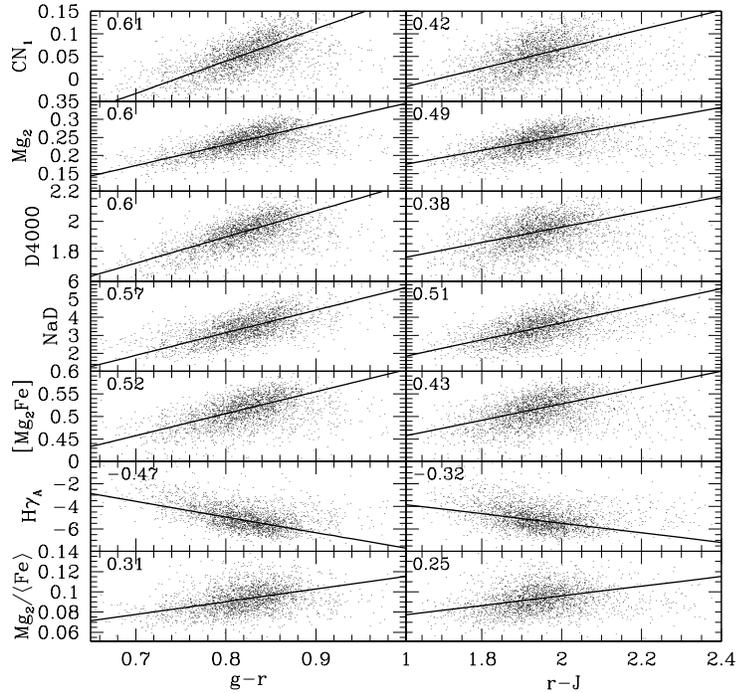}}
\caption{Spectral absorption features (corrected for velocity
dispersion) of the galaxies in our sample plotted as a function of
$g-r$ and $r-J$ colours. Here we show only a subset of the 29
spectral indices analyzed in the paper. The indices are ordered
according to the significance of the correlation (given by the
correlation coefficient in each panel) with $g-r$.}\label{fig6}
\end{figure*}

\begin{figure*}
\centerline{\includegraphics[width=10truecm]{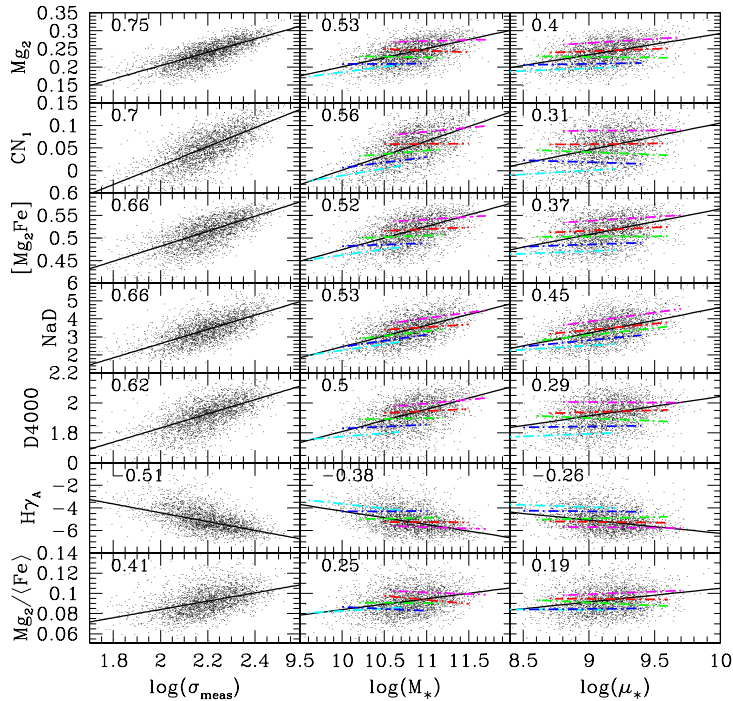}}
\caption{Spectral absorption features (corrected for velocity
dispersion) of the galaxies in our sample plotted as a function of
the measured velocity dispersion, the stellar mass and the stellar
surface mass density. The same subset of spectral features of
Fig.~\ref{fig6} is shown here.The indices are ordered according to
the significance of the correlation with velocity
dispersion. In the middle and right panels, the dot-dashed lines
show the linear relations for galaxies with similar velocity
dispersion (cyan: $\log\sigma_{meas}<2$, blue: $2\leq
\log\sigma_{meas}<2.1$, green: $2.1\leq \log\sigma_{meas}<2.2$, red:
$2.2\leq \log\sigma_{meas}<2.3$, magenta: $\log\sigma_{meas}\geq2.3$).}\label{fig7}
\end{figure*}

In Table 2 we summarize the results for a variety of different
Lick indices and for $g-r$ and $r-J$ colours. As discussed by
\cite{Thomas03a}, the $\alpha$-element to Fe abundance ratio can
be empirically quantified by the ratio between Mg and Fe indices.
Here we use both the $\rm Mg_2/\langle Fe\rangle$ and the $\rm
Mgb/\langle Fe\rangle$ index ratio.\footnote{The $\langle
Fe\rangle$ is the average of Fe5270 and Fe5335 index strengths.}
It is also possible to define other combinations of indices that
are almost independent of $\alpha$-enhancement. Those investigated
here are $\rm [MgFe]^\prime$, as defined by \cite{Thomas03a}, and
$\rm [Mg_1Fe]$ and $\rm [Mg_2Fe]$, as given by BC03. For each
index, we list in Table 2 the slope of the relation (columns 3 and
7), the dispersion about the relation (columns 4 and 8), the
significance of the relation according to a Spearman rank-order
test (columns 5 and 9) and the variation in index strength over
the colour interval that contains 90 percent of the points
($\Delta I$), normalized to the dispersion in the fitted relation
(columns 6 and 10). There is a small fraction of outliers in $g-r$
and $r-J$ colours (probably galaxies with large systematic
uncertainties in their photometry) which significantly affect the
fits. To account for this, we decide to exclude those galaxies
that lie above 2.5 times the dispersion in the colour-magnitude
relations shown in Figs.~\ref{fig2} and~\ref{fig4}. In this way,
about 12 percent of the galaxies are excluded from the fit. We
than adopt the same robust technique to fit relations between
index strengths and colours. We consider as statistically not
significant those relations with a correlation coefficient lower
than 0.3. In Table 2 we also provide information about the
indices, including the mean signal-to-noise ratio of the index
(column 2), the elements that are expected to contribute most to
the index strength (column 11)  and the sensitivity of the index
to $\alpha$/Fe ratio (column 12). This information was taken from
\cite{henry99,Thomas03a} and \cite{thomas04}. In Fig.~\ref{fig6},
we illustrate a selection of (statistically significant) relations
between index strength and $g-r$, $r-J$ colours. Indices are
ordered from top to bottom with decreasing correlation coefficient
(indicated in each panel) with respect to $g-r$.

In Table 2, the quantities of most interest are the {\em relative} strengths of
the correlations between index strength and colour for different indices (as given
by the correlation coefficient and, e.g., $\Delta I$/disp). For $g-r$, the
strongest correlations are found for  the Mg- and CN-features, the 4000\AA\ break
D$_n$(4000), the Balmer absorption lines (H$\gamma_A$, H$\delta_A$ and H$\beta$)
and the $\alpha$-enhancement independent indices [Mg$_2$Fe], $\rm [MgFe]^\prime$
and [Mg$_1$Fe].  When we comparing these correlations with those
involving $r-J$ we see that:\\
\begin{enumerate}
\item D$_n$(4000) and the Balmer absorption lines
correlate more strongly with the optical
colour than with the optical-infrared colour.
\item The $\alpha$/Fe independent
Mg-Fe composite indices correlate better with
optical-infrared than with optical colour.
\item Mg features and NaD correlate
equally well with both colours.
\item $\rm Mg_2/\langle Fe\rangle$ shows a
weak correlation with both colours.
\end{enumerate}
In summary,  both optical and optical-infrared colours
are sensitive to metallicity and to element abundance ratios. However,
the age of the stellar populations (as given by $\rm D_n(4000)$ and Balmer
absorption lines) influence the optical colours more
strongly than the optical-infrared colours. One caveat  is
that the colours are measured for the galaxy as a whole,
whereas the Lick indices
are measured within the 3 arcsecond fibre aperture. Gradients in index strength
may thus affect the analysis.
We have checked whether  our results change if we use
SDSS ``fibre'' colours (calculated within a 2.5 arcsec aperture) rather than the
Petrosian colours. There are changes in the strengths of some of the correlations,
but our main qualitative conclusions remain unchanged.

We now turn to the correlations between index strengths and
structural parameters, such as stellar mass ($\log M_\ast$),
measured velocity dispersion ($\log \sigma_{meas}$) and
surface mass density ($\log \mu_\ast$). For consistency with
the previous analysis, we apply here our robust fitting
technique only on galaxies within 2.5$\times$ $disp$ of the
$g-r,M_r$ and $r-J,M_r$ relations. The results of the fits
for all the indices are summarized in Table 3, where we give
slope, dispersion, correlation coefficient and $\Delta
I$/disp as before. Fig.~\ref{fig7} shows a selection of
absorption indices against velocity dispersion, stellar mass
and surface mass density, ordered with decreasing
correlation coefficient with respect to velocity dispersion.

The main result of this analysis is that all indices correlate
much more strongly with velocity dispersion than with any other
structural parameter. This is another strong indication that the
properties of elliptical galaxies depend most strongly on the
velocity dispersion of the system. We note that both the velocity
dispersion and the spectral indices are measured within the fibre,
while stellar mass and surface mass density pertain to the galaxy
as a whole. One might thus worry that this might artificially
strengthen the correlation between stellar absorption indices and
velocity dispersion. We have checked that similar results are
obtained when we correlate the absorption indices with the
inferred velocity dispersion obtained by diving the stellar mass
by the Petrosian half-light radius. We conclude that our main
result is robust against these effects.

Finally, we have checked whether there are residual correlations
between age-dependent and metallicity-dependent spectral indices
and structural parameters such as surface mass density. The
dot-dashed lines in the middle and right panels of Fig.~\ref{fig7}
show the linear relations with stellar mass and surface stellar
mass density, for galaxies in five bins of velocity dispersion.
This clearly shows that, after removing the dependence on velocity
dispersion, the absorption line strengths do not show any
significant residual correlation with stellar mass and, even more
so, with surface stellar mass density. Kennicutt's law of star
formation \citep{kennicutt98} relates the star formation rate in
galactic disks to the surface density of cold gas in the galaxy.
The fact that we see very little correlation between spectral
indices and surface density for ellipticals suggests that global
processes, related to the mass of the system, have been more
instrumental in setting the properties of galactic spheroids.

\section{Discussion}

The main aim of this paper is to investigate what can be learned
from the optical, optical-infrared and infrared colours of
elliptical galaxies. We have put together a sample of ellipticals
from a matched sample of galaxies with photometry from both SDSS
and 2MASS and spectroscopy from SDSS. We study the correlations
between colours and a wide variety of different structural
parameters, as well as the correlations between stellar absorption
features, colours and structural parameters. We find that:

\begin{enumerate}
\item  Luminous and massive elliptical  galaxies  have redder
optical, optical-infrared and infrared colours than less luminous
and lower mass ellipticals. \item The optical, optical-infrared
and infrared colours of elliptical galaxies correlate more
strongly with stellar mass and velocity dispersion than with other
structural parameters. The dispersion about these two relations is
also smaller. \item Both optical and optical-infrared colours are
sensitive to metallicity and to element abundance ratios. However,
the age of the stellar populations influences the optical colours
somewhat more strongly than the optical-infrared colours.
\end{enumerate}

\begin{figure}
\centerline{\includegraphics[width=10truecm]{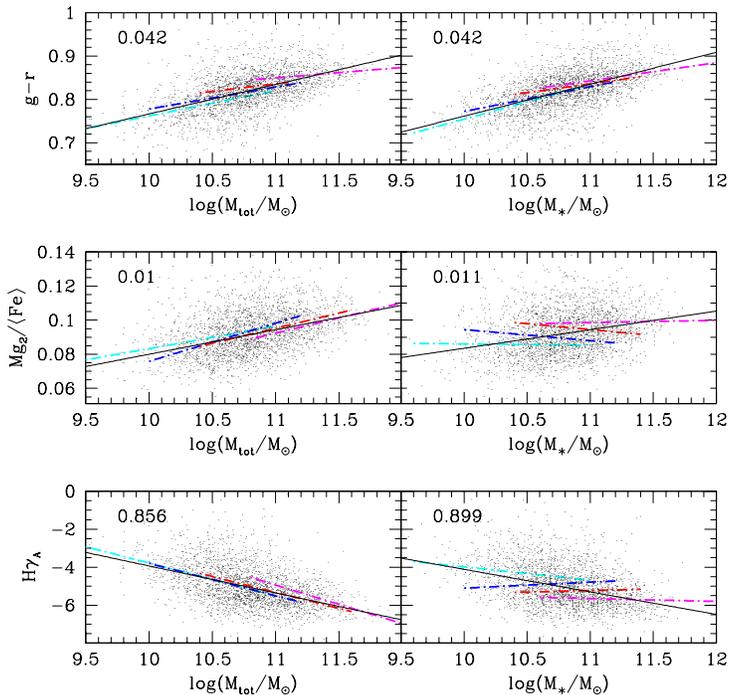}}
\caption{\it Left: $g-r$ colour, $\rm Mg_2/\langle Fe\rangle$ and
$\rm H\gamma_A$ index strengths against total mass $M_{tot} =
(2\sigma_{meas})^2 R_{50,z}/G$. The black solid line in each panel
represents the linear relation fitted for the sample as a whole,
while the coloured, dot-dashed lines show the relations for
galaxies in four bins of stellar mass (cyan: $\log M_\ast<10.5$,
blue: $10.5\leq \log M_\ast<10.8$, red: $10.8\leq \log
M_\ast<11.2$, magenta: $\log M_\ast \geq 11.2$). Right: $g-r$
colour, $\rm Mg_2/\langle Fe\rangle$ and $\rm H\gamma_A$ index
strengths against stellar mass. The coloured, dot-dashed lines
show here the linear relations for galaxies in four bins of total
mass (cyan: $\log M_{tot,inf}<10.5$, blue: $10.5\leq \log
M_{tot,inf}<10.8$, red: $10.8\leq \log M_{tot,inf}<11.2$, magenta:
$\log M_{tot,inf} \geq 11.2$). In each panel the rms scatter about
the relation is also indicated. }\label{fig8}
\end{figure}

Moreover, our results show that all Lick indices correlate more
strongly with velocity dispersion, either as measured within the
fibre or as quantified by the ratio between stellar mass and
Petrosian half-light radius, than with any other structural
parameter studied. As demonstrated by \cite {cappellari05}, the
velocity dispersion scaled by the size of the system (in this
case, the effective radius) provides a remarkably good estimate of
the total (stellar plus dark matter) mass of the galaxy. We have
also studied correlations between colours, spectral indices and
the mass inferred from the measured velocity dispersion $M_{tot} =
(2\sigma_{meas})^2 R_{50,z}/G$. In Fig.~\ref{fig8} we compare
these relations with those against stellar mass for $g-r$ colour,
a representative age-sensitive index ($H\gamma_A$) and a
representative metallicity-sensitive index ($\rm Mg_2/\langle
Fe\rangle$). In the left-hand panels the dot-dashed lines show the
relations with total mass at fixed stellar mass, while in the
right-hand panels they show the relations with stellar mass at
fixed total mass. The middle and bottom panels show that the
relation with total mass for galaxies with similar stellar mass
has almost the same slope as the relation obtained for the sample
as a whole. On the contrary, at fixed total mass, the relation
with stellar mass becomes much shallower.

However, the dependence on these two physical parameters is less
distinguishable in the case of $g-r$ colour. It is difficult to
tell from Fig.~\ref{fig8} which is the most fundamental parameter
between stellar and total mass. Since both the absorption features
and the velocity dispersion (which enters in the estimate of total
mass) are measured within the fiber, while stellar mass is a
global quantity, one might worry that the relation with total mass
is strengthened by aperture effects. We checked the correlation
between colours, spectral features and stellar mass in the fiber $
M_{*,fiber}$ (obtained by scaling stellar mass $M_\ast$ with the
ratio between fiber and Petrosian z-band luminosity) and show part
of the results in Table ~\ref{mass1}. It is shown that both the
colours and spectral indices correlate more strongly with the mass
inferred from the velocity dispersion than with the stellar mass
measured within the fiber in the sense that the dispersion with
the total mass is smaller that that of stellar mass in the fiber
and $\rm \Delta I/disp$ in Col. (4) is larger than that in Col.
(7).

In summary, only from the results of this paper, it is difficult
to draw a robust conclusion whether the stellar mass or the total
mass is the more fundamental parameter for the elliptical
galaxies. But our results do suggest that the star formation
history of nearby elliptical galaxy is primarily determined by the
mass of the system.

\begin{table*}
\caption{Correlations between colours, spectral indices and the
mass inferred from the measured velocity dispersion $M_{tot,inf} =
(2\sigma_{meas})^2 R_{50,z}/G $ and stellar mass contained in the
fiber $ M_{*,fiber}$.  Columns 2,5 give the slope of the relation;
columns 3,6 give the dispersions about the relation; columns 4,7
give the change in colours (or index strength) along the fitted
relation over the interval in structural parameter that contains
90 percent of the points, normalized to the dispersion about the
fitted relation.}\label{mass1} \centering
\begin{tabular}{|l|rrr|rrr|}
\noalign{\smallskip} \hline \noalign{\smallskip} colour (or Index)
& \multicolumn{3}{c}{$\log M_{tot}/M_{\odot}$} &
\multicolumn{3}{c}{$\log M_{*,fiber}/M_{\odot}$} \\
 & slope & disp & $\rm \Delta I/disp$ & slope & disp &
 $\rm \Delta I/disp$ \\
 (1) & (2) & (3) & (4) & (5) & (6) & (7) \\
\noalign{\smallskip} \hline \noalign{\smallskip}
(g-r)  & 0.0652 & 0.0294 &  2.5996 & 0.0732 & 0.0301 &  2.4904 \\
(r-i)  & 0.0333 & 0.0202 &  1.9403 & 0.0368 & 0.0206 & 1.8343 \\
(r-J)  & 0.1242 & 0.0740 &  1.9675 & 0.1292 & 0.0762 & 1.7352 \\
(r-K)  & 0.1700 & 0.1031 &  1.9349 & 0.1970 & 0.1045 & 1.9308 \\
$Mg_1$           & 0.0440  & 0.0161 & 3.275  &  0.0429 & 0.0184 &  2.412 \\
$H\delta_A$          & -1.2735 & 0.9328 & -1.636 & -1.1952 & 0.9586 & -1.289 \\
$D4000$                  & 0.1765  & 0.0793 & 2.667  &  0.1839 & 0.0837 &  2.273 \\
$[MgFe]^\prime$          & 0.4691  & 0.2812 & 1.999  &  0.4649 & 0.2946 &  1.632 \\
$[Mg_1Fe]$           &  0.0465 & 0.0216 & 2.58   &  0.0471 & 0.0227 &  2.146 \\
$[Mg_2Fe]$           &  0.0567 & 0.0250 & 2.718  &  0.0561 & 0.0269 &  2.157 \\
$Mg_2/\langle Fe\rangle$ & 0.0144  & 0.0108 & 1.598  &  0.0125 & 0.0115 &  1.124 \\
$Mgb/\langle Fe\rangle$  & 0.1457  & 0.2035 & 0.858  &  0.1158 & 0.2075 &  0.577 \\
\noalign{\smallskip} \hline \noalign{\smallskip}
\end{tabular}
\end{table*}

Our analysis of spectral absorption features has shown that both
optical and optical-infrared colours are sensitive to variations
in total stellar metallicity and, to a lesser extent, element
abundance ratios. Age has a
more significant effect on the optical colours than on the
optical-infrared colours of ellipticals.
The fact that optical-infrared colours are as strongly
correlated with velocity dispersion as the optical colours
argues for the fact that metallicity (and possibly element
abundance ratios) is the {\em primary} driver of the
colour trends among
elliptical galaxies. This is in agreement with  the results by
\cite{gallazzi05}, who show that in contrast
to metallicity,  the luminosity-weighted ages of
early-type galaxies in SDSS are only weakly dependent on mass.

\section*{Acknowledgments}
R.C. acknowledges the financial support of MPG for a
visit to MPA and expresses her gratitude for the hospitality
during her visit at MPA. This project is partly supported by NSFC
10173017, 10403008, 10133020, 10073016, NKBRSF 1999075404,
Shanghai Municipal Science and Technology Commission No.
04dz05905.

A.G. and S.C. thank the Alexander von Humboldt
Foundation, the Federal Ministry of Education and Research, and
the Programme for Investment in the Future (ZIP) of the German
Government for their support. We thank the anonymous referee for
his/her helpful suggestions to greatly improve this paper.

Funding for the creation and distribution of the SDSS Archive has been provided
by the Alfred P. Sloan Foundation, the Participating Institutions, the National
Aeronautics and Space Administration, the National Science Foundation, the U.S.
Department of Energy, the Japanese Monbukagakusho, and the Max Planck Society.
The SDSS Web site is http://www.sdss.org/.

The SDSS is managed by the Astrophysical Research Consortium (ARC) for the
Participating Institutions. The Participating Institutions are The University
of Chicago, Fermilab, the Institute for Advanced Study, the Japan Participation
Group, The Johns Hopkins University, Los Alamos National Laboratory, the
Max-Planck-Institute for Astronomy (MPIA), the Max-Planck-Institute for
Astrophysics (MPA), New Mexico State University, University of Pittsburgh,
Princeton University, the United States Naval Observatory, and the University
of Washington.


\label{lastpage}
\end{document}